# Mind the gap between theory and experiment

Andrei Kiselev, Jeonghyeon Kim, and Olivier J. F. Martin

**Abstract** After briefly introducing the surface integral equation method for the numerical solution of Maxwell's equations, we discuss some examples where numerical simulations based on effectively fabricated nanostructures can provide additional insights into an experiment. Focusing on plasmonics, we study Fano resonant systems for optical trapping, realistic dipole antennas for near-field enhancement, and hybrid nanostructures that combine plasmonic metals with dielectrics refractive index sensing. For those systems, the same experimental detail can play a very different role, depending on the type of physical observable. For example, roughness can significantly influence the near-field, but be totally unnoticed in the far-field. It can affect molecules adsorbed on the surface, while refractive index sensing can be fully immune to such roughness. Approaching the experimental situation as closely as possible is certainly a challenging task and we demonstrate a simple approach based on SEM images for that. Altogether, bridging the gap between theory and experiment is not such a trivial task. However, some of the simple steps illustrated in this chapter can help build numerical models that match the experiment better.

## 1 Introduction

We did not have the pleasure to meet Werner S. Weiglhofer and only know of him through his scientific publications. In spite of his too short career, they are extremely numerous, diverse and impactful. Following the Web of Science categories, one notices that these contributions do not only cover optics and electromagnetics, but also reach out to applied physics, materials sciences and – of course – mathematics. They are very well cited; his works on demystified negative index of refraction [1], and that on light-propagation in helicoidal bianisotropic media [2], at the top of his list of citations. Working at the Department of Mathematics of the University

Nanophotonics and Metrology Laboratory, Swiss Federal Institute of Technology Lausanne (EPFL), EPFL-STI-NAM, Station 11, CH-1015 Lausanne, e-mail: olivier.martin@epfl.ch

of Glasgow, it is not surprising that Werner's publications have a strong theoretical flavour and have inspired many theoretical works. Yet, analysing their citations further indicates that these theoretical developments inspired numerous experimental projects. As an example, among the citations of his work on light-propagation in helicoidal bianisotropic media [2], a third of the citing articles report experiments. This illustrates how well Werner succeeded in bridging the gap between theory and experiments! Obviously, theory is very important and progress within the realm of theoretical physics is often fascinating in itself. Evidently, new theories are often the driver behind new experimental work. This chapter, however, focuses on the inverse process, where experimental work requires numerical support as close as possible to the experimental situation. After briefly presenting the numerical technique we have developed for over a decade to solve Maxwell's equations, we discuss three different experimental situations where we attempted to model the real experiment as closely as possible.

## 2 Computational electromagnetics

At the onset of studying a given experimental situation lies the fundamental question of the choice of the most appropriate numerical method. It is fair to say that there is not one single numerical technique that is fit for all situations and even for the narrow field of plasmonics, which is the focus of our work, numerous approaches exist as illustrated in a recent review article [3]. Furthermore, each numerical method can be put to good use as long as it is utilized wisely and carefully. Especially, sufficient efforts must be undertaken to characterize the algorithm beforehand, to make sure that it will converge well for the problem at hand and is free from spurious behaviours. This task is especially thankless, but of paramount importance if the numerical results are to be trusted. Note that it does not only apply to home-developed numerical codes, but should be equally undertaken with commercial packages that should never be trusted blindly, even if they produce beautiful and colourful images!

To assess the accuracy of a numerical technique and obtain a metric to quantify it, one usually resorts to canonical problems. Unfortunately, there are essentially only two such problems for which a reference solution exists (the quasi analytical Mie solution): the scattering by a sphere for three-dimensional (3D) problems or by a cylinder for two-dimensional (2D) geometries [5]. Figure 1 illustrates this approach for a 2D solution obtained with a volumetric Green's tensor approach [4]. In this case, two different incident polarizations must be considered, with the electric field either perpendicular to the cylinder axis (transverse electric or TE field) or parallel to the cylinder axis (transverse magnetic or TM field). The differential cross section can be computed as a function of the scattering angle and compared with the quasi analytical Mie solution, Fig. 1(a). This panel indicates that many features exist in that response, which need to be reproduced accurately with the numerical method. A more quantitative metric is obtained by integrating the difference between this cross section and the Mie solution over all scattering angles and repeating the calculations with an increasing number of discretized elements, Fig. 1(b). In principle, the error should decrease as the number of elements increases. However, this behaviour is far from monotonous since it includes different facets of the numerical problem: on the one side, a finer mesh approximates the scatterer better and should provide a more accurate solution; on the other hand, it requires a larger numerical matrix to be solved, which is more difficult, especially when the matrix condition number increases, as is the case here [6,7]. Consequently, plateaus appear in the convergence curve, Fig. 1(b). We also notice that the polarization influences the solution accuracy, reminiscent that in electromagnetics all field components do not behave in the same way: some are continuous across materials' boundaries, others are not [8].

Experimental situations are usually much more complicated than a sphere or a cylinder and we will show in Sec. 3.2 that it is possible to use reciprocity to assess the accuracy of numerical results produced for complex geometries.

In this chapter, we focus on the surface integral equation (SIE) method for the numerical solution of Maxwell's equations. An interesting feature of such a formulation is that the boundary conditions at the edge of the computation window are included in the equations and need not be taken care of by using *ad hoc* prescriptions, such as absorbing boundary conditions or perfectly matched layers [9]. Indeed, these boundary conditions are already included in the kernel of the integral equation, and can take different forms, like infinite homogeneous space [10], surfaces or stratified media [11, 12], or waveguide cavities [13]. There is of course a price to pay for this: except for infinite homogeneous space where the kernel is known analytically [10], it must be evaluated numerically, usually by resorting to plane waves or eigenmodes expansions [11, 14].

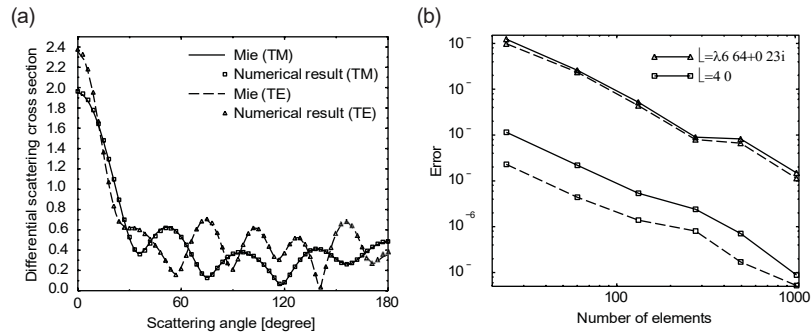

**Fig. 1** Measuring the numerical solution accuracy for light scattering by an infinite cylinder in vacuum illuminated by a plane wave normal to the cylinder. Two polarizations are considered: transverse electric (TE) with the electric field normal to the cylinder axis and transverse magnetic (TM) with the electric field along the cylinder axis. (a) Comparison between the numerical solution and the reference Mie solution for a dielectric cylinder with relative permittivity $\varepsilon = 4$ and a size parameter $x = \pi \sqrt{\varepsilon} d / \lambda = 10.43$, where $d$ is the cylinder diameter, $\varepsilon$ its relative permittivity and $\lambda$ the wavelength in vacuum. (b) Relative error between the Mie and numerical solutions (defined as the square of the difference between the numerical and Mie far-field amplitudes, normalized to the square of the analytical amplitude) as a function of the number of discretized elements for two different materials $\varepsilon$, for TE (solid lines) and TM (dashed lines) polarizations. Adapted from Ref. 4 with permission, copyright IEEE 2000.

The SIE is constructed from the combination of an equation for the electric field and one for the magnetic field; different weighted combinations can be used here [15]. The volume integral form of Maxwell's equations is transformed into a surface equation using Gauss' theorem and the solution is computed from unknown electric and magnetic currents defined only on the surface of the scatterer [16]. This is very advantageous since only that surface needs to be discretized; on the other hand, a limitation of this approach is that the resulting matrix is dense since each mesh is connected to all the other meshes in the system, different, e.g., from the finite difference time domain method, where only nearest neighbours are connected [17]. The resulting system of linear equations is constructed through a Galerkin procedure, where Rao-Wilton-Glisson functions are used both as basis and test functions [18]. The accuracy of the method strongly depends on the order used for the quadrature in the Galerkin scheme [19]. Once the surface currents are known, different observables can be computed, from the near-field to the different cross-sections [20], or even the force and torque produced by the incident light on the nanostructure [21, 22].

## 3 Approaching experimental situations

Having settled for the numerical technique, we wish to address the question of modelling the geometry of a real experiment as accurately as possible and will do that in the context of three different plasmonic systems. This field of research studies the interaction of light with coinage metals, like gold, silver, aluminium or heavily doped semiconductors [23]. When light impinges on a nanostructure made from such a metal, it resonantly excites the free electrons in the metal, producing a very strong near-field at the vicinity of the nanostructure [24, 25]. It is quite remarkable that nanostructures much smaller than the wavelength can exhibit such strong resonances; the reason being the localisation of the free charges in a specific pattern associated with each optical resonance [26, 27].

### 3.1 Fano-resonant systems

In principle, any resonant system has an optical response with a Lorentzian shape [5]. This is also true for a plasmonic nanostructure, as long as only one single resonance is excited like in a small particle or a dipole antenna [28]. On the other hand, as soon as more than one resonances are present, the lineshape can become very complicated with several different peaks. A prominent family of such irregular responses is the so-called Fano lineshape, following the name of Ugo Fano who discovered them while interpreting atomic spectroscopy experiments [29]. In the context of plasmonics, Fano resonances occur when two modes are present in the system, often a bright mode (a mode that radiates into the near-field, like a dipole) and a dark mode (a mode that does not radiate into the far-field, like a quadrupole) [30]. In plasmonics, the intrinsic losses associated with the metal make the resonances relatively broad [31], such that several modes can overlap and interact, even when their exact resonance frequencies are different. The bright mode is excited by the incoming excitation and produces some near-field that can in turn excite the dark mode [32]. The latter will also produce a near-field that affects the bright mode. Depending whether both responses are in- or out-of-phase (i.e., depending on the excitation wavelength), they will interfere constructively or destructively, producing the asymmetric lineshape.

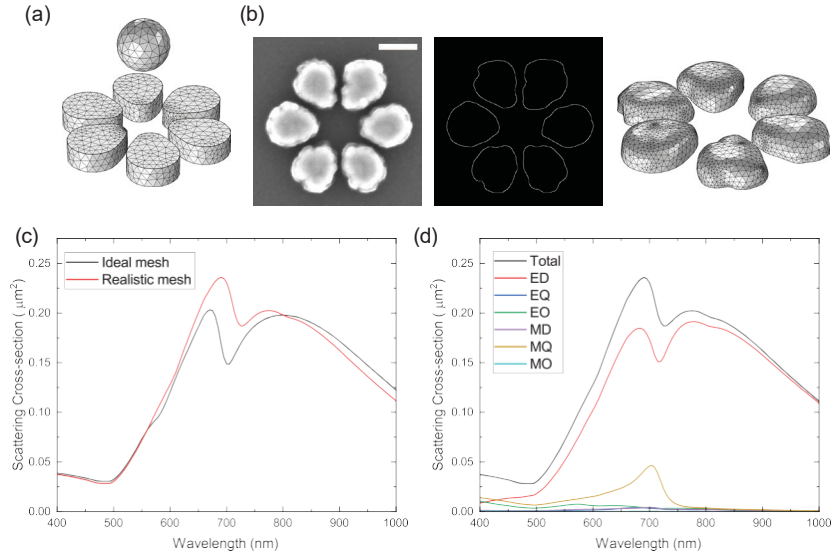

**Fig. 2** Optical trapping with an hexamer. (a) Six gold nanostructures are deposited on a substrate and produce a strong optical field under linear polarized illumination, which can trap a gold nanosphere. The mesh for an ideal structure is shown. (b) (from left to right) Realization of a realistic mesh based on a SEM image of an effectively realized nanostructure (scalebar 100 nm), which outline is determined using the Canny edge detector from the Scikit-image Python package. The realistic mesh is built in Blender from this outline by inspection and comparison with the SEM image. (c) Comparison of the scattering cross sections for the ideal and realistic meshes as a function of the wavelength in vacuum $\lambda$. (d) Multipolar decomposition of the scattering cross section obtained from the realistic mesh into Cartesian multipoles: electric/magnetic dipoles (ED/MD), electric/magnetic quadrupoles (EQ/MQ), and electric/magnetic octupoles (EO/MO).

The interest for plasmonic Fano resonant systems lies in the fact that they exhibit very narrow spectral features, in spite of the significant losses inherent to plasmonic metals. This is useful for sensing, where the quality factor (i.e., the resonance width) determines the sensitivity and several experiments have been performed along those lines [33–40].

Here, we are rather interested in the very strong optical near-field generated by Fano-resonant nanostructure, as illustrated in Fig. 2(a). Six gold nanostructures are positioned on a circular ring forming a hexamer and illuminated with linear polarized light. These structures produce a strong near-field gradient that will exert a force on a nearby gold sphere (the experiment is performed in water), which will become trapped at the centre of the structure. Once the sphere is trapped, the structure becomes a heptamer and its spectrum changes. In order to guide this experiment, it is important to have and accurate description of its spectral response, especially to choose the best excitation wavelength. To this end, the experimentally realized hexamer is used to build a finite elements mesh for the SIE calculations, Fig. 2(b): first, the edge of the structure is obtained from the scanning electron microscope (SEM) image using the edge detector in the Scikit-image Python package with a standard deviation $\sigma$ = 7 pixels, which corresponds to the spatial extent for that edge detection [41]. This 2D outline is imported into Blender vers. 3.2 [42], and extruded to form a 3D object, whose sharp edges are smoothed using the bevel function, and sculpted until it mimics the shape inferred from the SEM image. A more accurate approach would consist in using tomography in an electron microscope [43], e.g., high-angle annular dark-field scanning transmission electron microscopy, which provides amazing 3D reconstructions of plasmonic nanostructures [44, 45].

Inspecting the scattering cross sections (SCSs, which correspond to the power flow scattered by the structure in the far-field) for the ideal and realistic meshes in Fig. 2(c) we observe some differences in the Fano lineshapes, especially the magnitudes of the different peaks. Based on the realistic meshes, the SCS can be decomposed into different Cartesian multipoles [46]. Interestingly, although the electric dipole dominates the response of the system, we also observe a rather important magnetic quadrupole around $\lambda$ = 700 nm, Fig. 2(d). The Fano resonance proceeds from the interaction between the electric dipole (bright mode) and the magnetic quadrupole (dark mode).

## 3.2 Near-field of plasmonic antennae

The near-field distribution at the vicinity of plasmonic nanostructures is the driver for all interactions with molecular or atomic systems, such as fluorescence [47–51], or surface enhanced Raman spectroscopy (SERS) [52–54]. Computing an absolute value of this field enhancement is an extremely difficult task, which certainly still deserves important research developments. In this section, we wish to show that the exact nanostructure geometry can play a significant role for the computed near-field. To this end, we consider a dipole antenna made from gold with two 100 nm long arms separated with a 25 nm gap. It is possible to fabricate such a nanostructure fairly accurately with a high resolution electron beam system and ion etching [55]. However, it can happen that the produced nanostructure resembles more a pair of potatoes than two perfect parallelepipeds, as shown in the SEM image in Fig. 3(a). Using a similar approach as that described in Sec. 3.1, it is possible to infer from the SEM image a finite element mesh for that structure and use it to compute the optical response of the realistic particle. Interestingly, in the far-field both the perfect rectangular and the realistic one have the same spectral response with a resonance at $\lambda$ = 630 nm and a modest magnitude difference. Considering both antennae as electromagnetic objects and drawing the electric field lines, produces also quite similar impressions, with maybe slightly denser field lines for the real antenna, Fig. 3(c).

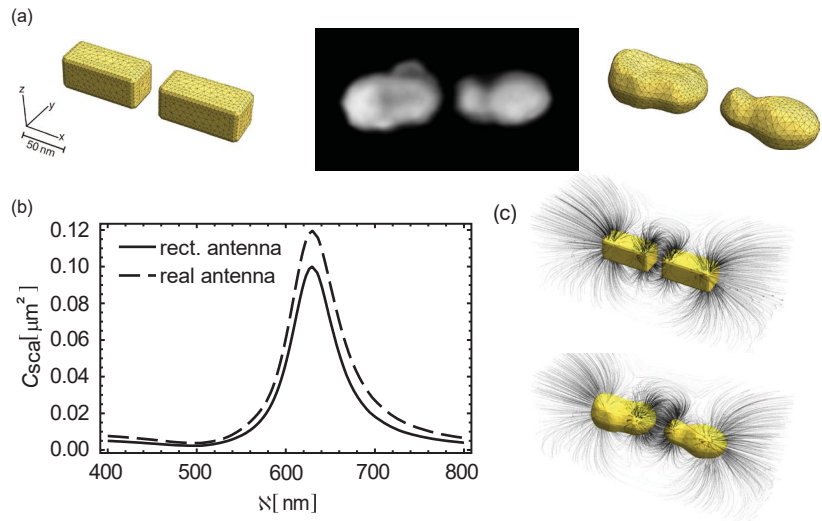

**Fig. 3** Gold plasmonic dipole antenna built from two 100 nm long gold nanostructures (40× 40 nm² cross section) with a 25 nm gap. (a) (left to right) Mesh used for the ideal rectangular structure, SEM image of an effectively realized structure and mesh used for the realistic structure. (b) Scattering cross sections computed for the ideal and realistic antennae. (c) Electric field lines derived from the electromagnetic models. Panels (a) and (b) adapted from Ref. 56 with permission, copyright ACS 2011.

A very different behaviour is observed in the near-field of both antennae, as shown in Fig. 4, where we compute the near field intensity enhancement around the left arm of each antenna at the resonance wavelength $\lambda$ = 630 nm. Let us first focus on the field at the close vicinity of the metal (2 nm), unwrapped like for a geography map. The incident field has unit intensity and the intensity is enhanced by three orders of magnitude for both antennae. The near-field distribution is very different for each geometry, with most of the enhancement at the vicinity of the corners for the ideal structure and a broader and smoother field distribution for the realistic structure. Using the ideal geometry to compute how molecules would be driven by that antenna produces very different results compared to those obtained with the realistic structure. Especially those "hot-spots" at the antenna corners are very unlikely to appear in an experiment, where fabricated metal nanostructures always exhibit significant roughness. We will come back to this issue of roughness in Sec. 3.3 and show that its importance strongly depends on the physical situation at hand. At larger distances from the surface, 10 nm in Fig. 4, both field distributions become much more similar, without any noticeable "hot-spots" near the rectangular corners. The resemblance of a numerical model with the experimental reality is therefore especially important in the ultimate near-field, the region where, for example, molecules interacting with the structure, would be located. Sadly, many simple numerical models used to study near-field interactions with plasmonic nanostructures rely on ideal, parallelepiped structures.

The previous observations on the near-field distribution at the vicinity of a plasmonic nanostructure prompt us to make a short discussion on reciprocity and its use to validate numerical models. Reciprocity is a complicated concept that can be easily misused and we refer the interested reader to the excellent review article by Caloz et al. where it is discussed in details [57]. Briefly, in an optical experiment with a source and a detector, reciprocity requires that the system responds in a similar way when the source and the detector are exchanged. This can be illustrated with the plasmonic antennas considered in this section. Figure 5 shows the field enhancement for the ideal dipole antenna (top row) and the realistic antenna (bottom row). The solid lines show the enhancement of the light radiated by a dipole source detected in the far-field at $(x;y;z) = (0;0;106\text{ nm})$. Three different dipole locations indicated by the black dots are considered, as well as two different dipole orientations: parallel to the antenna axis (red lines) and normal to the antenna axis (green lines). When the dipole is in the gap of the antenna, the intensity enhancement resembles the scattering cross sections shown in Fig. 3(b), with again a slightly larger enhancement for the realistic antenna. This coupling between the dipole and the antenna strongly depends on the dipole orientation and no field enhancement is observed when the dipole is normal to the antenna axis (green lines). As soon as the dipole is displaced away from the antenna centre, the enhancement decreases significantly (note the different vertical axis ranges for the second and third columns in Fig. 5). In this case, the ideal antenna still exhibits a Lorentzian response with a single spectral feature, indicating that this geometry supports only one electric dipole resonance. This is not the case for the realistic geometry, where several Cartesian multipoles interact to produce a more complicated response. These data indicate that molecules spin-coated on a dipole antenna will experience very different radiation enhancements, depending on their location on the nanostructure, with the molecules located close to the gap benefiting most from that enhancement.

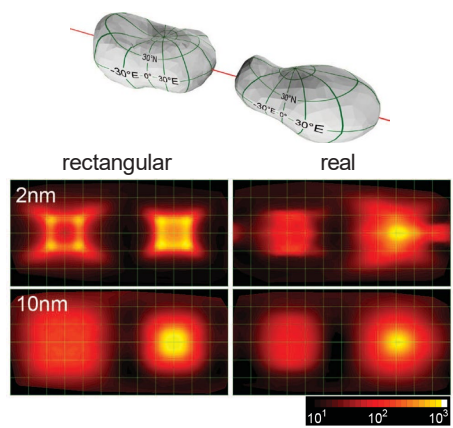

**Fig. 4** Comparison of the near-field intensity enhancement computed around the ideal rectangular antenna and the real one deduced from the SEM image. The equirectangular projection coordinate system used to map the field intensity is shown at the top for the real antenna. The intensity maps are shown for the left arm of each antenna (rectangular or real) at two different separation distances from the metal: 2 nm and 10 nm. Adapted from Ref. 56 with permission, copyright ACS 2011.

The crosses and dots shown in Fig. 5 are the results of separate calculations where a dipolar source was used in the far-field at the location $(x;y;z) = (0;0;106$ nm$)$ and the field intensity was computed near the antenna at the location marked with the black dot. It should be noted that reciprocity applies to the field components, not the total intensity. Hence, for the first calculations with the dipolar source close to the antenna, only the $x$-polarization of the far-field was computed; for the second calculations, the dipole located in the far-field was oriented in the same $x$-direction, while the intensity of only the $x$- (respectively $z$-) component of the electric field at the black dot was computed for the red (respectively green) lines. Altogether, this procedure corresponds to exchanging the source and the detector and the perfect agreement between the lines and the symbols in Fig. 5 indicates that those numerical results fulfil reciprocity.

This provides a useful way of checking the accuracy of numerical results beyond the comparison with a reference solution on a very simple, canonical, geometry discussed in Sec. 2. This check is very easy to perform when the numerical method at hand can handle infinite geometries, as is the case for algorithms based on the integral form of Maxwell's equations.

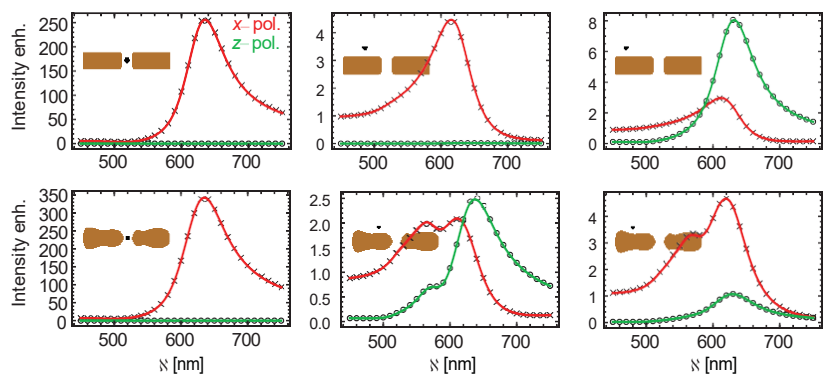

**Fig. 5** Solid lines show the enhancement of the intensity radiated to the far-field by a dipole source in the proximity of an ideal rectangular (top row) or realistic (bottom row) nanoantenna. The source is located in the $x$-$z$-plane (same axes as in Fig. 3 as indicated by the dots in the insets, oriented in $x$-direction along the antenna axis (red curves) or in $z$-direction normal to the antenna axis (green curves) while far-field detection is at $(x;y;z)=(0;0;106$ nm$)$ and in $x$-polarization. Symbols represent the reverse process with source and detection points exchanged: the $x$-polarized source is located at $(x;y;z) = (0;0;106$ nm$)$ and the detection is in the near-field in $x$-polarization (crosses) or $z$-polarization (circles). Adapted from Ref. 56 with permission, copyright ACS 2011.

To conclude this section, let us note that in Fig. 4 we dared to compute the electromagnetic field at a very short distance from the metal: 2 nm. Whether a pure classical electromagnetic approach is sufficient to do so is of course an intricate question. There appears however to be a consensus that down to that distance, it is still reasonable to do so: for shorter distances, one should resort to more sophisticated models, beginning first by including non-local effects [58], while even shorter distances require more complicated approaches such as time dependent density functional theories [59] or advanced quantum (or quantum-corrected) models [60–62].

### 3.3 Hybrid nanostructures

So far, we have considered the strong optical resonances that can be excited in plasmonic metals and are associated with their very high density of free electrons [23]. Surprisingly, strong resonances can also be observed in high refractive index dielectric nanostructures by virtue of so-called Mie resonances [63]. In short, one can say that plasmonic nanostructures have essentially a fundamental resonance with an electric dipolar character, while dielectric structures have a magnetic dipolar fundamental resonance. An interesting question is whether combining these two materials can open up a new field of investigations where electric and magnetic

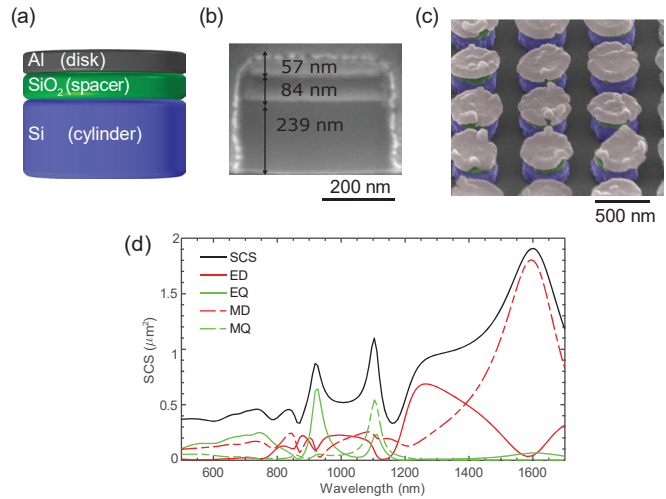

**Fig. 6** (a) Hybrid nanostructure built from a 220 nm thick Si cylinder with a 60 nm thick Al disk, separated with a 75 nm thick $SiO_2$ spacer. (b) SEM image of the effectively fabricated nanostructure. (c) Falsed colours SEM image of nanostructures with recessed $SiO_2$ spacer to improve the sensitivity. (d) Cartesian multipoles decomposition of the scattering cross section (SCS) in electric/magnetic dipoles (ED/MD) and electric/magnetic quadrupoles (EQ/MQ).

resonances are combined? Such hybrid nanostructures are beginning to emerge, with only a few experimental demonstrations so far [64, 65], including one from our group [66]. Theoretical studies have demonstrated that indeed coupling metals and dielectrics produce a rich spectrum with some very narrow features like anapoles [67].

Figure 6(a) shows the geometry for such a hybrid nanostructure we realized for sensing applications [66]. Nominally, it is composed of a 220 nm thick Si cylinder base with a 60 nm thick Al disk cap, separated with a 75 nm $SiO_2$ spacer; the overall structure has a diameter of 470 nm. The dielectric spacer thickness can be adjusted to control the coupling between the modes in the dielectric Si cylinder and those in the Al plasmonic disk, providing control on the spectral response of the system, as studied in detail in Ref. 68 and illustrated in the calculations shown in Fig. 6(c). Note that due to absorption of Si, this structure's response is mainly in the near infrared range of the optical spectrum. Note also that the fundamental mode of the structure, around $\lambda$ = 1600 nm is magnetic dipolar.

The effectively fabricated nanostructure realized by reactive ion etching, shown in Fig. 6(b), has dimensions that are quite close to the ideal structure with a 239 nm thick Si cylinder, a 84 nm spacer and a 57 nm thick Al disk. Their utilization for bulk refractive index sensing was tested experimentally, yielding a rather modest sensitivity of 208 nm/RIU [66], about half the best value obtained for bulk refractive index sensing with pure plasmonic nanostructures. The reason for that disappointing sensitivity can be well understood from numerical simulations. It is known that the spatial overlap between the near-field produced by the sensing nanostructure and the analyte is key for the sensitivity [69–71]. Unfortunately, the field distribution computed for the ideal structure, shown in Fig. 7(a) indicates that most of the electric field remains within the dielectric spacer. This observation prompted the idea to etch away some of this spacer, sufficiently to expel some of the electric field into the background, but not too much to compromise the nanostructure stability. This can be performed with an additional wet etch step with hydrofluoric acid [68]. The resulting nanostructures are shown with false colours SEM images in Fig. 7(c) with the Si cylinder in blue, the recessed $SiO_2$ spacer in green and the Al disk in grey. Figure 7(b) and (c) indicate that this treatment increases indeed the electric field in the background. Experimentally, the sensitivity increased from 208 nm/RIE to 245 nm/RIE [68].

Further insights into this modest experimental improvement are provided in Fig. 8(a), which shows the maximum of the near-field enhancement computed in the background for the different distributions shown in Fig. 7. First, we notice that the strongest enhancement is obtained for only lightly etched spacers and located close to the sharp metal edge in the metal (see Fig. 7(b)); we are again facing an issue related to sharp unrealistic geometrical features as in Sec. 3.2. Furthermore, there is a wavelength shift for the main resonance as the spacer diameter decreases, Fig. 8(a). In that context, it is interesting to notice that the second resonance around $\lambda$ = 1100 nm is quite prominent for the more realistic nanostructures; to the extent that the field enhancement is almost as strong as that provided by the fundamental resonance at $\lambda$ = 1600 nm. This observation has important implications for the experiment and it would have been unnoticed if more realistic nanostructures had not been simulated. Figure 8(b) indicates that, on the other hand, the scattering cross section of the nanostructure is not very sensitive to the exact geometry, as was already the case in Sec. 3.2.

Finally, the fabricated hybrid nanostructures have an extremely rough Al top surface caused by the morphological growth of this metal, Fig. 6(c). Even with advanced nanofabrication techniques, such roughness cannot be avoided [72], and an interesting question is whether it disturbs the nanostructure optical response. Since the SIE relies on a triangular mesh, it is possible to build models for rough nanostructures, as shown in Figs. 7(d) and (f). This roughness was simply created by adding Gaussian noise to the flat original surface, with a maximum amplitude of 8 nm for the rough surface and 60 nm for the very rough one. From the field distributions in

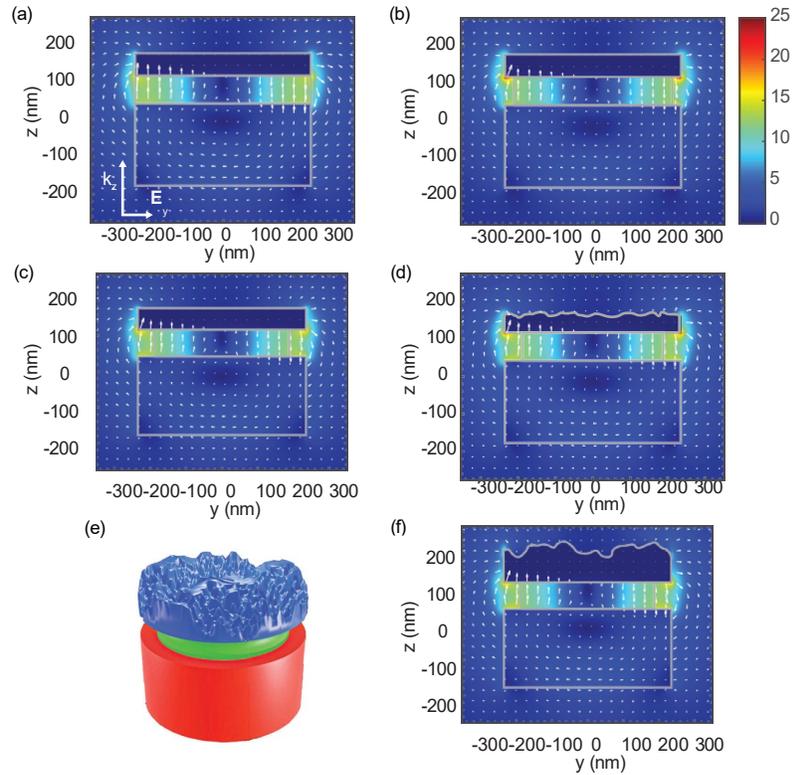

**Fig. 7** The electric field amplitude enhancement colormaps for different hybrid nanostructures, together with the vector electric field distribution. (a) Complete ideal structure, (b) ideal structure with SiO$_2$ spacer reduced to 90%, (c) respectively. Structures with the SiO$_2$ spacer reduced to 80% and a rough cap Al disk obtain by Gaussian noise with maximum amplitude (d) 8 nm and (f) 60 nm. (e) Sketch of the model computed in panel (f) with a recessed spacer and a rough Al surface. All the structures are illuminated from the bottom as indicated in panel (a) and immersed in air.

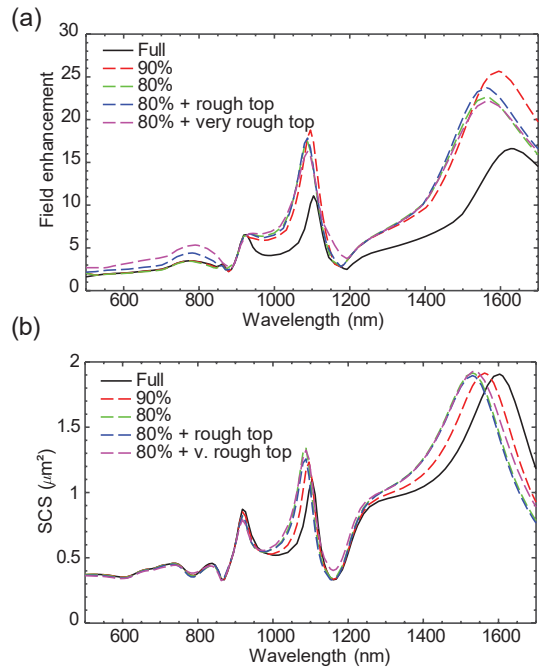

**Fig. 8** (a) Maximum field amplitude enhancement in the background for the distributions shown in Fig. 7 and (b) corresponding scattering cross sections (SCSs). The full structure and those where the SiO$_2$ spacer has shrunk to 90% and 80%, as well as the rough and very rough Al disks are considered.

Fig. 7, one does not notice significant differences between the perfectly flat surfaces and the roughened ones. This is confirmed by the maximum field enhancement and the scattering cross sections in Fig. 8. Especially for the latter, which does not change at all compared with the perfectly flat surface. This information is useful to guide experimental efforts in the direction where they really impact the optical response: investing time in developing a process that would produce flat Al surfaces is not worth for those experiments. Actually, the physical reason for this lack of influence of the roughness is in the polarization used for those measurements: the incident field being parallel to the top Al surface, it does not feel a roughness that is essentially normal to that surface.

## 4 Summary and outlook

In summary, we have discussed some examples where numerical simulations based on effectively fabricated nanostructures can provide additional insights into an experiment. While calculations are often used at the inception of a project, closing the loop and redoing calculations from the experimental data is very rewarding and one should probably perform the full cycle "simulations → nanofabrication → characterization → measurements → simulations" several times to gain additional insights into the underlying phenomena.

For the specific case of plasmonics considered here, we have noticed that the same experimental detail can play a very different role, depending on the type of physical observable. For example, roughness can significantly influence the near-field, but be totally unnoticed in the far-field. It can affect molecules adsorbed on the surface, while refractive index sensing can be fully immune to such roughness. As scientists, we often have a taste for "perfection" and wish to fabricate nanostructures that have ideal shapes. Obviously, this is not possible since materials have their own minds and will not submit to our square-headed epitome. Knowing how far to go (or not) towards perfection can save a lot of time and efforts.

Approaching the experimental situation as closely as possible is certainly a challenging task. An important issue is to build numerical models that mimic the effectively realized nanostructures. Here, we have used a very simple approach based on SEM images. Tomography in an electron microscope provides a more sophisticated way forward to retrieve accurate 3D representations of nanostructures [43–45]. At the same time, it is clear that each individual nanostructure will be different and efforts should also be invested in building statistics to determine the details that really matter for the optical response.

Another key issue that we have not addressed in this chapter is the dielectric function used for the simulations. Even for plasmonic metals, one finds many different values in the literature, which can produce quite different optical responses. In addition, it is very unlikely that a metal deposited in a specific machine will exactly match those values from the literature. In principle, one should characterize each material with ellipsometry to retrieve its exact dielectric function, which is quite tedious... and might not even provide a more accurate solution: ellipsometry requires thick metal films (at least 100 nm thick), while plasmonic nanostructures are often much thinner and have a different roughness, which can influence at least the absorption.

Altogether, bridging the gap between theory and experiment is not such a trivial task. However, some of the simple steps illustrated in this chapter can help build numerical models that match the experiment better. In any case, the very first step in that endeavour is to check the convergence and stability of the numerical method at hand.

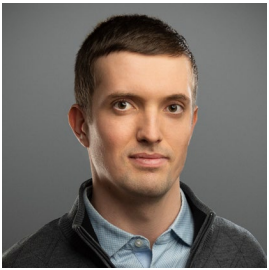

**Andrei Kiselev** Andrei Kiselev is a 4th year PhD student in the Laboratory of Nanophotonics and Metrology of the Swiss Federal Institute of Technology in Lausanne (EPFL, Switzerland). Andrei defended his bachelor and master theses at Lomonosov Moscow State University, where his research was focused on photonic crystals and metamaterials based on bulk Dirac semimetals. Andrei has interests in BioNanophotonics, a field he was studying during his internship in the laboratory of Prof. Aleksandra Radenovic with the goal of developing a portable DNA sequencer. Current Andrei's field of research focuses on the exploration of optical interactions of nanoparticles with the goal to control the optical forces between them and eventually become able to create a laser driven nanofactory-on-chip capable of assembling nanostructures on demand.

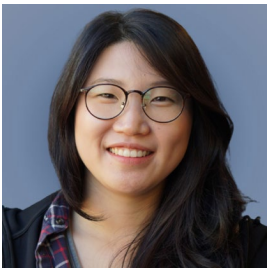

**Jeonghyeon (Jenna) Kim** is a Ph.D. student in Photonics at the École polytechnique fédérale de Lausanne (EPFL). She received a bachelor's degree in Electrical Engineering and a master's degree in Integrated Technology from Yonsei Univer-sity, South Korea. She joined the Nanophotonics and Metrology Laboratory (NAM) at EPFL in 2017, where she is carrying her Ph.D. thesis on optical trapping of gold nanoparticles. Her current research interests include nanophotonics, optical manipulation, colloids, statistical physics, and interface science.

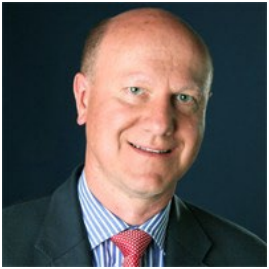

**Olivier J.F. Martin** studied physics at the Swiss Federal Institute of Technology Lausanne (EPFL) and conducted his PhD at IBM Zurich Research Laboratory, where he studied semiconductor physics. After a stay at the University of California in San Diego, he became Assistant Professor at the Swiss Federal Institute of Technology Zurich (ETHZ). In 2003 he was appointed at the EPFL, where he is currently Full Professor of Nanophotonics and Optical Signal Processing, Director of the Nanophotonics and Metrology Laboratory. Between 2005 and 2017 he was Director of the EPFL Doctoral Program in Photonics (approx. 100 PhD students) and between 2016 and 2020 he was Director of the EPFL Microengineering Section (1'000 students). In that latter capacity he has conducted an in-depth reform of the study plan and introduced the new EPFL Master in Robotics. Dr. Martin conducts a comprehensive research that combines the development of numerical techniques for the solution of Maxwell's equations with advanced nanofabrication and experiments on plasmonic systems. Applications of his research include optical antennas, metasurfaces, nonlinear optics, optical nano-manipulations, heterogeneous catalysis, security features and optical forces at the nanoscale. Dr. Martin has authored over 300 journal articles and holds a handful of patents and invention disclosures. He received in 2016 an Advanced Grant from the European Research Council on the utilization of plasmonic forces to fabricate nanostructures; he is a Fellow of the Optical Society of America and Associate Editor of Advanced Photonics and of Frontiers in Physics.

18    Andrei Kiselev, Jeonghyeon Kim, and Olivier J. F. Martin